# *In vivo* ultrasound-switchable fluorescence imaging

**Tingfeng Yao, [1,2]\* Shuai Yu, [1,2]\* Yang Liu[1,2] and Baohong Yuan[1,2+]**

[1]Ultrasound and Optical Imaging Laboratory, Department of Bioengineering, University of Texas at Arlington, Arlington, TX 76010, USA

[2]Joint Biomedical Engineering Program, University of Texas at Arlington and University of Texas Southwestern Medical Center at Dallas, Dallas, TX 75235, USA

[+]Correspondence and requests for materials should be addressed to B.Y. (baohong@uta.edu)

\* These authors contributed equally to this work.

## Abstract

Ultrasound-switchable fluorescence (USF) imaging was recently developed to overcome the limitation of the poor spatial resolution of the conventional fluorescence imaging in centimeter-deep tissue. However, *in vivo* USF imaging has not been achieved so far because of the lack of stable near-infrared contrast agents in a biological environment and the lack of data about their biodistributions. In this study, for the first time, we achieved *in vivo* USF imaging successfully in mice with high resolution. USF imaging in porcine heart tissue and mouse breast tumor via local injections were studied and demonstrated. *In vivo* and *ex vivo* USF imaging of the mouse spleen via intravenous injections was also successfully achieved. The results showed that the USF contrast agent adopted in this study was very stable in a biological environment, and it was mainly accumulated into the spleen of the mice. All the results of USF imaging were validated by a commercial micro-CT.

## Introduction

Biomedical optical imaging based on fluorescence has been intensively developed during the past decades and is playing important roles in many applications, such as fluorescence microscopy[1-2], two-dimensional (2D) fluorescence planar imaging[3-4], and three-dimensional (3D) fluorescence diffuse optical tomography[5-6]. Besides fluorescence microscopy, imaging based on fluorescence contrast in deep tissues (such as centimeters) attracts a lot of interests too because it has many unique advantages compared with other modalities, such as x-ray, ultrasound, magnetic resonance imaging (MRI), and positron-emission tomography (PET). For example, it can provide unique opportunities to simultaneously image multiple biological targets via multiple colors and therefore make it possible to investigate the interactions among these targets[7-8]. This may include examples such as monitoring migration of cancer cells or cell clusters into blood vessels in metastasis in deep tissues by differentiating the cells from the vessels via labeling them with different colored fluorophores. Also, it may be used to increase the imaging specificity to a biological target in deep tissues (such as cancer stem cells or cell clusters) by simultaneously identifying the multiple biomarkers of the target via multicolored fluorophores.

However, fluorescence imaging in deep tissues has been facing several dilemmas since its invention because of the tissue's high scattering property[9]. First, its spatial resolution significantly degrades with the increase of tissue thickness, such as from submicrons for microscopy in a tissue sample with a thickness of a few hundred microns to a few millimeters in a tissue with a thickness of a few centimeters. The low resolution will lead to unwanted spatial overlap of different fluorescence signals and difficulty differentiating them. Second, to achieve high spatial resolution in deep tissue, the sensitivity can be an issue since the detectable photons become limited due to the improvement of spatial resolution. Third, nonspecific photons, which may come from tissue autofluorescence, laser leakage, or stray light in the environment, are always noises and reduce the specificity and sensitivity of signal to its contrast agents in deep tissues.

To address these challenges, several technologies have been or are being investigated, such as ultrasound-modulated fluorescence[10-11] or luminescence[12], ultrasound-induced temperature-controlled fluorescence[13-15] or luminescence[16], and time-reversed ultrasonically encoded optical focusing[17-21]. These technologies have been shown to be able to achieve high acoustic resolution and retain fluorescence (or luminescence) contrast in deep scattering media.

During the past years, we have developed a new technology, ultrasound-switchable fluorescence (USF) imaging. What makes USF different from others is the unique contrast agent that can be switched on via ultrasound-induced thermal energy (or temperature rise) only in the ultrasound focal volume[15, 22-25]. By scanning the ultrasound focus, a 3D USF image can be acquired with fluorescence contrast and ultrasound resolution. Unlike x-ray based computed tomography (CT), MRI, ultrasound and photoacoustic imaging[26], USF signal has extremely high specificity to its contrast agent, which means the detected USF signal can only come from its contrast agent. This feature is similar to PET imaging in which the detected γ photons only come from the positrons emitted by the injected radioactive isotope[27]. In addition, USF signal strength and dynamic pattern can be externally manipulated by controlling the ultrasound exposure, which makes the USF signal uniquely differentiated from the noise to achieve high signal-to-noise ratio (SNR) or sensitivity[23-24]. Besides structural, functional and molecular imaging, USF can be seamlessly developed for high-intensity focused ultrasound (HIFU) treatment as a tool predicting pre-treatment, monitoring intro-treatment and evaluating post-treatment.

Although the results are promising from the technologies mentioned above, all of them are still in their early or concept development stage, and all the results are demonstrated in tissue-mimic phantoms (such as highly optical scattering intralipid solution or gels) or tissue samples (such as chicken breast tissue or porcine muscle tissues). Significant technical challenges hinder their use in real *in vivo* applications. To achieve those potential biomedical applications via fluorescence contrast, it is important to demonstrate the *in vivo* imaging feasibility and validate the accuracy. In this study, overcoming the technical barriers, we, for the first time, successfully demonstrate *in vivo* USF imaging in mice and validate the results by micro-CT imaging. The success of *in vivo* USF imaging is an important step to push this technology for future applications.

## Results

**Quantification of the effect of the driving voltage on the spatial resolution of USF imaging in a silicone tube-based tissue phantom.** To quantify the effect of the driving voltage of HIFU on the spatial resolution of the frequency-domain (FD)-USF system, we carried out USF imaging using a silicone tube-based tissue phantom. As shown in Figure 1(a), a silicone tube (with an inner diameter of 0.31 mm and outer diameter of 0.64 mm) was inserted into a piece of porcine muscle tissue (with a thickness of ~10 mm) at a depth of ~5 mm to simulate a blood vessel. Before USF imaging, the silicone tube was filled with indocyanine green (ICG)-encapsulated poly(N-isopropylacrylamide) (PNIPAM) nanoparticles (ICG-NPs) with a lower critical solution temperature (LCST) of ~24–25 °C (See Figure S1 in Supplementary Information (SI) for details). To obtain the lateral and axial one-dimensional profiles of the tube, a HIFU transducer was used to scan the tube laterally (with a step size of 127 μm and a scanning range of 6.604 mm) and axially (with a step size of 635 μm and a range of 10.16 mm).

Figure 1(b) shows the lateral (i.e., Y direction) one-dimensional USF profiles of the silicone tube with two different HIFU driving voltages of 80 mV and 140 mV from the function generator (i.e., peak-to-peak, Vpp; see the details about the system in Figure S9 in SI). The peak values of the USF profiles for the two driving voltages of 80 mV and 140 mV are about 267 mV and 626 mV, respectively. The higher driving voltage induces stronger USF signal strength because of the higher HIFU-induced temperature rise. As shown in Figure 1(c), the full width at half maximum (FWHM) of the lateral profiles for the two driving voltages are 1.08 mm and 1.39 mm, respectively. Figure 1(d) shows that the lateral FWHM of the USF profile of the tube increases with the rising of the HIFU driving voltage (60, 80, 100, 120, 140, 160, and 180 mV and the estimated ultrasound power: 0.43, 0.77, 1.21, 1.74, 2.36, 3.09, 3.90 W). This is because a higher driving voltage can induce a larger thermal focal volume, switch on more contrast agents, and lead to a stronger USF signal and a larger FWHM (i.e., a lower spatial resolution of USF imaging).

As shown in Figure 1(e), the peak values of the axial USF profiles (i.e., Z direction) for the driving voltage of 80 mv and 140 mV are about 207 mV and 661 mV, respectively. As shown in Figure 1(f), the FWHM of axial profiles for the two driving voltages are 3.31 mm and 5.37 mm, respectively. Figure 1(g) shows that the axial FWHM of the USF profile of the silicone tube increases with the rising of the HIFU driving voltage. It should be noticed that the axial FWHM is much larger than the lateral FWHM, which is due to the focal zone shape of the HIFU transducer. The lateral and axial FWHMs of the acoustic intensity focus was measured as 0.55 mm and 2.8 mm[28], respectively.

Figure (h) shows the lateral profile of the silicone tube with a driving voltage of 180 mV, and Figure (i) is the corresponding normalized USF signals as a function of time at different positions of the tube (right). It should be noted that the shapes of the USF signals are different when the HIFU is scanning across the tube (i.e., along the Y direction). The USF signal at the edge of the tube decays more slowly than that at the center of the tube. Due to the thermal diffusion, the USF signal will have a longer duration time than in the focus. Three reference signals chosen for correlation in image processing are based on this observation (see more details in SI Figure S5).

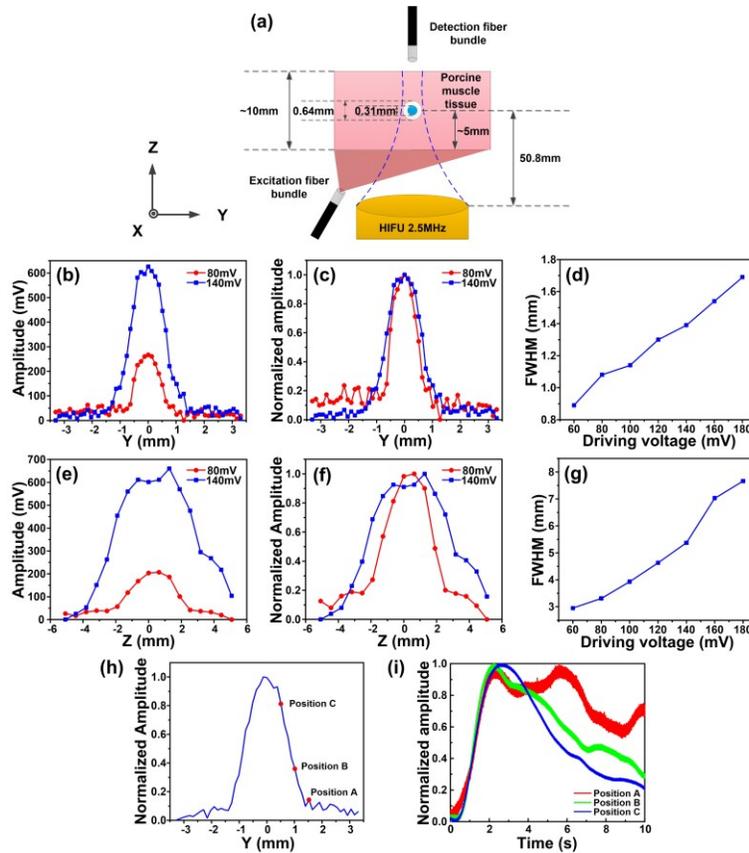

**Figure 1.** (a) The sample configuration, including the porcine muscle tissue, the silicone tube, the excitation and detection fiber bundles, and the HIFU transducer. (b) and (c) are the original and normalized lateral (i.e., Y direction) one-dimensional profiles of the silicone tube with a driving voltage of 80 mV (the red solid line with circles) and 140 mV (the blue solid line with squares), respectively. (d) The lateral FWHMs at different driving voltages. (e) and (f) are the original and normalized axial (i.e., Z direction) one-dimensional profiles of the silicone tube with a driving voltage of 80 mV (the red solid line with circles) and 140 mV (the blue solid line with squares), respectively. (g) The axial FWHMs at different driving voltages. (h) The lateral one-dimensional profile of the silicone tube with a driving voltage of 180 mV, and (i) the corresponding USF signals as a function of time at the position A (the red line), the position B (the green line) and the position C (the blue line).

**USF imaging of the contrast agent distribution in the porcine heart tissue via a local injection.** Figure 2(a) is a white-light photo of the porcine heart tissue (with a thickness of 10 mm). The USF contrast agent (37.5 µL) was mixed with the contrast agent of the CT (12.5 µL), and the mixed solution was locally injected into the heart tissue to form a single spot. Figure 2(b) shows the 2D and normalized fluorescence intensity distribution on the tissue top surface (i.e., on the XY plane). The image was acquired via an electron-multiplying charge-coupled-device (EMCCD) camera, the excitation light has a central wavelength of 808 nm, and the emission detection is >830 nm. The red box indicates the scan area of the USF imaging on the horizontal plane. A volume of 9.144 × 9.144 × 7.620 $mm^3$ was raster scanned by the HIFU transducer with a driving voltage of 120 mV. The lateral step size was 0.762 mm, and the axial step size was 1.524 mm. Figures 2(c1−c6) show the 2D USF images of the contrast agent distribution on the XY plane at different depths (i.e., Z direction). Figures 2(d1−d3) show the top view (XY plane), right view (YZ plane), and front view (XZ plane) of the 3D CT images of the CT contrast agent distribution, respectively. Similarly, Figures 2(e1−e3) show the top view, right view, and front view of the 3D USF images of the USF contrast agent distribution, respectively. Figure 2(f1) shows the co-registered 3D image of the two modalities, and Figures 2(f2−f5) are the 2D cross-section images on the XY plane at different depths. The green volumes or areas indicate the distribution of the USF contrast agent only, and the blue ones show the distribution of the CT contrast agent distribution only. The red ones represent the contrast agent distributions detected by both imaging modalities.

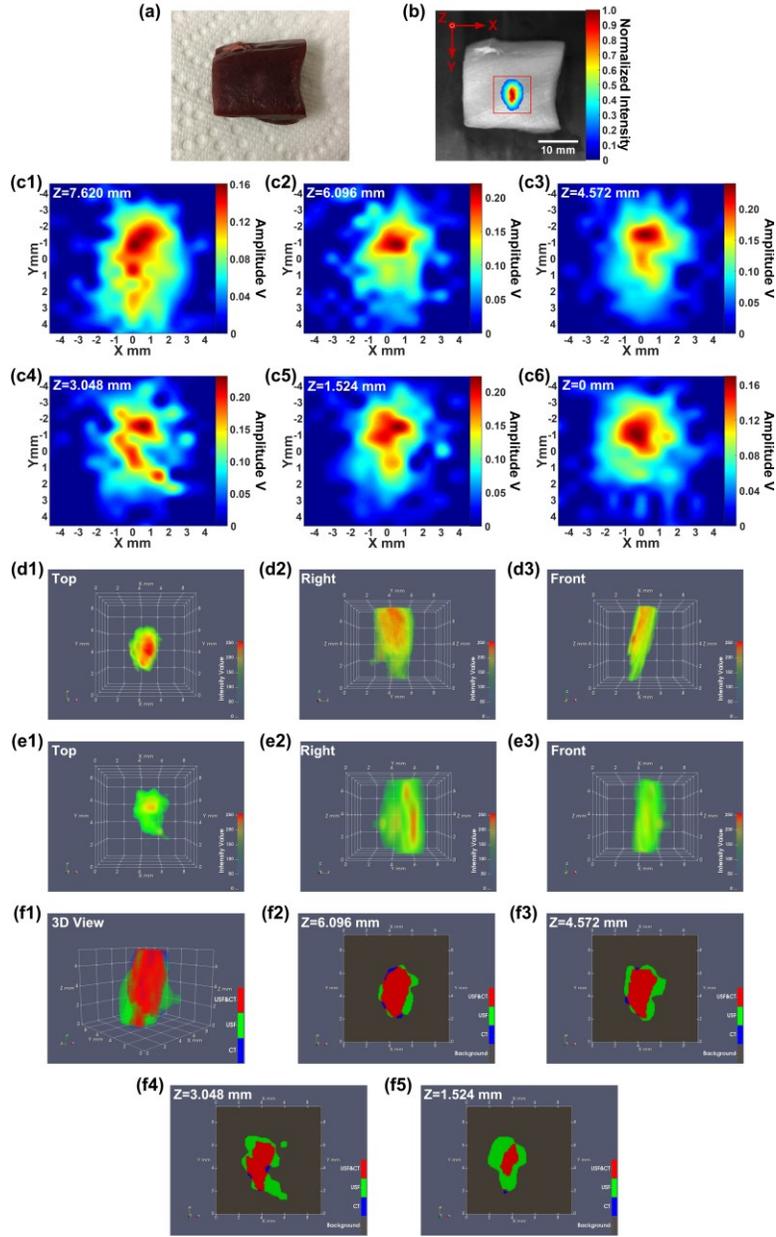

**Figure 2.** (a) A white-light photo of the porcine heart tissue. (b) The fluorescence image (Ex/Em=808/830 nm) of the mixed contrast agents distribution in the tissue. The red square indicates the scan area of the USF imaging on the horizontal (XY) plane. (c1–c6) 2D USF images on XY plane at different depths. (d1–d3) The top view (XY), right view (YZ), and front view (XZ) of the reconstructed 3D CT images. (e1–e3) The top view (XY), right view (YZ), and front view (XZ) of the 3D USF images. (f1) The 3D view of the co-registered image of the two modalities, and (f2–f5) the 2D cross-section images on XY plane at different depths.

***In vivo* USF imaging of the contrast agent distribution in a breast tumor on a mouse via a local injection.** Figure 3(a) shows the 2D fluorescence planar image (Ex/Em:808/830 nm) of the injected contrast agents in a mouse's breast tumor, and a white-light photo of the mouse is shown on the top right corner. The red box overlaid on the fluorescence image indicates the scan area of the USF imaging on the horizontal XY plane. A volume of 7.112 × 7.112 × 7.620 mm³ was raster scanned by the HIFU transducer with a driving voltage of 250 mV. The lateral step size was 0.508 mm, and the axial step size was 2.54 mm. Figure 3(b) shows the shell temperature of the mouse acquired using an infrared camera (see the details in Methods). Because the shell temperature of the mouse tumor is ~35 °C, the LCST of the USF contrast agent used for this *in vivo* experiment is ~35–36 °C rather than ~24–25 °C, which is suitable for phantom experiments at room temperature. Figures 3(c1–c4) show the

2D USF images of the contrast agent distribution on the XY plane at different depths (i.e., Z direction). Figures 3(d1–d3) show the top view (XY), right view (YZ), and front view (XZ) of the 3D CT images of the CT contrast agent distribution, respectively. Similarly, Figures 3(e1–e3) show the top view, right view, and front view of the 3D USF images of the USF contrast agent distribution, respectively. Figure 3(f1) shows the 3D co-registered USF and CT image, and Figures 3(f2–f5) are the cross-section images on the XY plane at different depths.

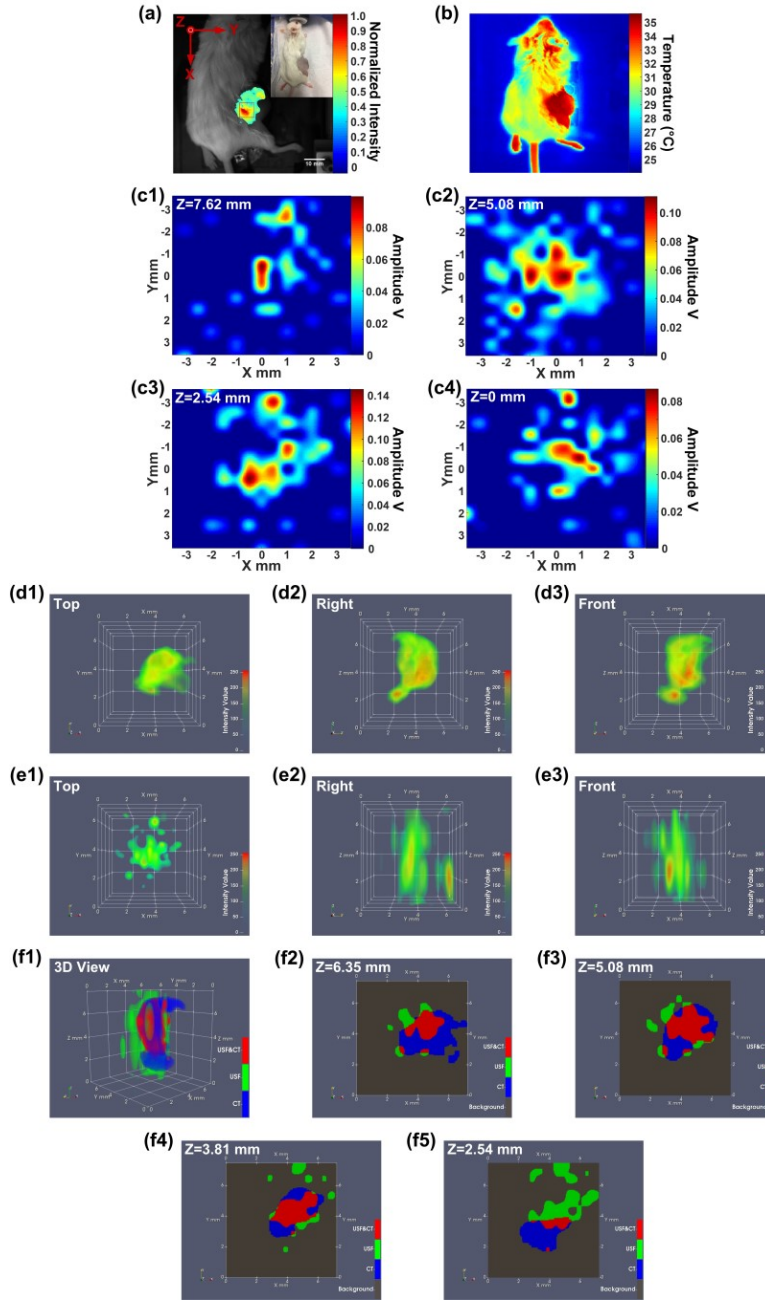

**Figure 3.** (a) A white-light photo of the mouse with a breast tumor (top-right) and the fluorescence image (Ex/Em=808/830 nm) of the mixed contrast agents locally injected in the tumor. The USF scan area is indicated via a red square on the horizontal XY plane. (b) An image of the shell temperature of the mouse acquired via an infrared camera. (c1–c4) 2D USF images on XY plane at different depths. (d1–d3) The top view (XY), right view (YZ), and front view (XZ) of the reconstructed 3D CT images. (e1–e3) The top view (XY), right view (YZ), and front view (XZ) of the 3D USF images. (f1) The 3D view of the co-registered image of the two imaging modalities, and (f2–f5) the 2D cross-section images on XY plane at different depths.

***In vivo*** **USF imaging of the contrast agent distribution in a mouse's spleen via an intravenous injection.** Figures 4(a1–a3) show the fluorescence images (Ex/Em:808/830 nm) of the mouse body at 0, 3 and 9 hours after the intravenous (i.v.) injection of 150 µL mixed contrast agent via the tail vein (100 µL ICG-NPs with an LCST of ~35–36 °C as USF contrast agent and 50 µL ExiTron nano 12000 as CT contrast agent). The ICG-NPs were found to be accumulated in the region of the mouse spleen and the fluorescence signal reached to a peak at ~3–4 hours after the i.v. injection (see the details of ICG-NPs metabolic process in SI). The red box on Figure 4 (a3) indicates the USF scan area on the horizontal XY plane. A volume of 9.144 × 12.192 × 6.096 $mm^3$ was raster scanned by the HIFU transducer with a driving voltage of 200 mV. The lateral step size was 0.762 mm, and the axial step size was 2.032 mm. Figures 4(b1–b4) show the 2D USF images of the contrast agent distribution on the XY plane at different depths (i.e., Z direction). Figures 4(c1–c3) show the top view (XY), right view (YZ), and front view (XZ) of the 3D CT images of the CT contrast agent distribution, respectively. The white dashed lines represent the contours of these spleen CT images. Similarly, Figures 4(d1–d3) show the top view, right view, and front view of the 3D USF images of the USF contrast agent distribution, respectively. For direct comparison, the same white dashed lines as shown in Figures 4(c1–c3) indicating the contours of the spleen CT images were also overlaid on these USF images. Figure 4(e1) shows the 3D co-registered USF and CT image, and Figures 4(e2–e5) are the cross-section images on XY plane at different depths.

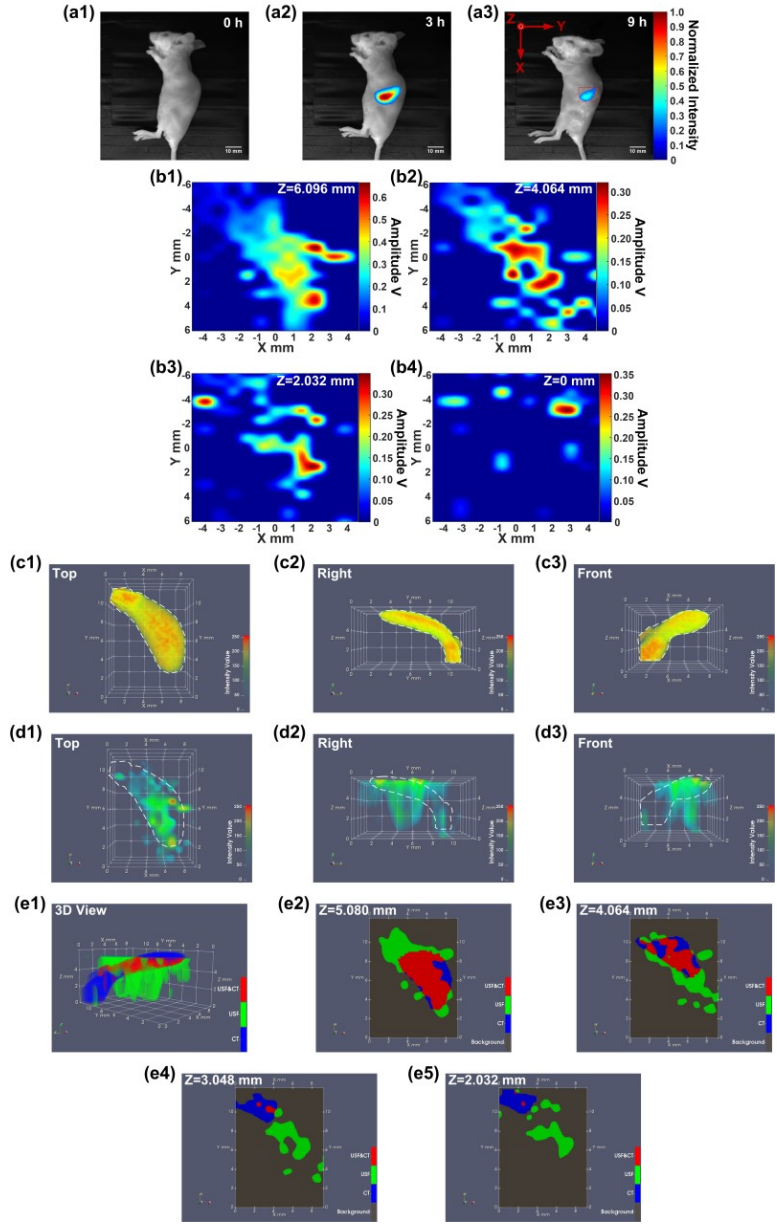

**Figure 4.** (a1–a3) 2D planar fluorescence images of the mouse at different times after i.v. injection of the mixed contrast agent (Ex/Em=808/830 nm). The red box on the Figure (a3) indicates the USF scan area on the horizontal XY plane. (b1–b4) 2D USF images on XY plane at different depths. (c1–c3) The top view (XY), right view (YZ), and front view (XZ) of the reconstructed 3D CT images. (d1–d3) The top view (XY), right view (YZ), and front view (XZ) of the 3D USF images. (e1) The 3D view of the co-registered image of the two imaging modalities, and (e2–e5) the 2D cross-section images on XY plane at different depths.

***Ex vivo* imaging of the mouse's spleen.** To further validate the results of the *in vivo* USF imaging of the mouse spleen, *ex vivo* USF imaging of the mouse spleen was carried out. Figure 5(a) is a white-light photo of the major organs of the mouse dissected after the USF imaging. Figure 5(b) is the fluorescence image overlaid on the black-and-white image of the organs. As shown in Figure 5(b), the fluorescence image (Ex/Em:808/830 nm) of the organs confirmed that most of the ICG-NPs were accumulated into the spleen as it emitted the strongest fluorescence signal. Also, this figure indicates that the distribution of ICG-NPs in this spleen may not be uniform because the lower left part of the spleen on this image emits a stronger fluorescence signal than does the upper right part. The stomach, liver, and kidney emit relatively low fluorescence signals. The heart and intestine show no or very weak signals. Figure 5(c) shows the reconstructed CT image of the same organs. Similar to the fluorescence image, the spleen also shows the strongest CT signal, and the distribution of the CT contrast agent in the spleen is also not uniform. The liver and some parts of the large intestine show moderate CT signal, and the kidney, heart, stomach, and small intestine all show much weaker CT signals.

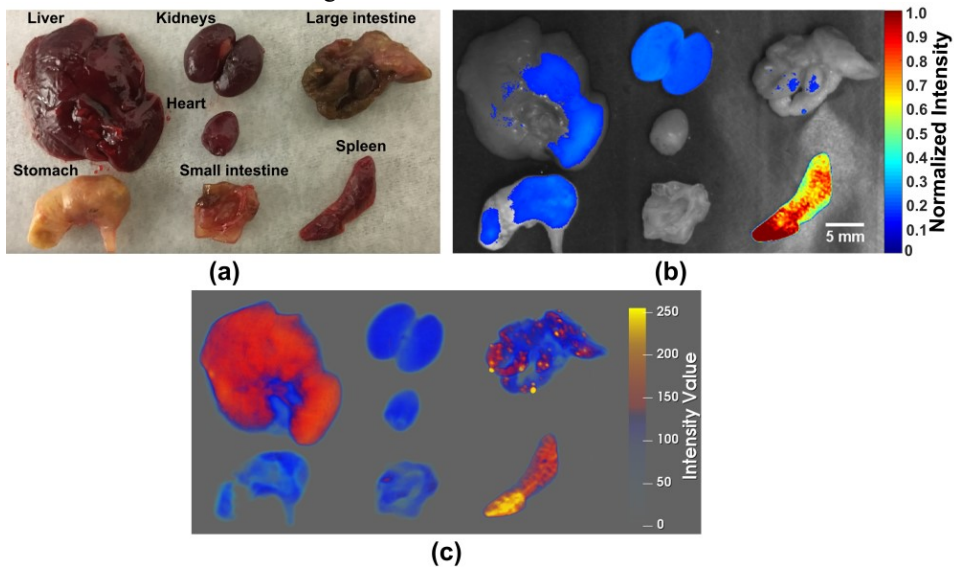

**Figure 5.** Bio-distribution of ICG-NPs (USF contrast agent) and ExiTron nano 12000 (CT contrast agent) in mice. (a) A white-light photo of the major organs dissected after USF imaging. (b) A 2D planar fluorescence image (Ex/Em=808/830 nm) of the organs. (c) A reconstructed CT image of the organs.

As shown in Figure 6(a), to simulate the situation of the spleen inside the mouse body, the spleen was inserted into a piece of porcine heart tissue (with a thickness of 1 cm) in which the spleen was close to the surface. Figure 6(b) shows the 2D fluorescence image of the tissue overlaid on a black-white background image. The red box indicates the scan area of the USF imaging on the horizontal XY plane. A volume of $13.716 \times 7.620 \times 8.128$ mm$^3$ was raster scanned by the HIFU transducer with a driving voltage of 140 mV. The lateral step size was 0.762 mm, and the axial step size was 2.032 mm. Figures 6(c1–c5) show the 2D USF images of the contrast agent distribution on the XY plane at different depths (i.e., Z direction). Figures 6(d1–d3) show the top view (XY), right view (YZ), and front view (XZ) of the reconstructed 3D CT images of the CT contrast agent distribution, respectively. The white dashed lines indicate the contours of the spleen CT images. Similarly, Figures 6(e1–e3) show the top view, right view, and front view of the 3D USF image of the USF contrast agent distribution, respectively. For direct comparison, the same white dashed lines as shown in Figures 6(d1–d3) indicating the contours of the spleen CT images were also overlaid on these USF images. Figure 6(f1) shows the 3D co-registered USF and CT image, and Figures 6(f2–f5) are the corresponding cross-section images on XY plane at different depths.

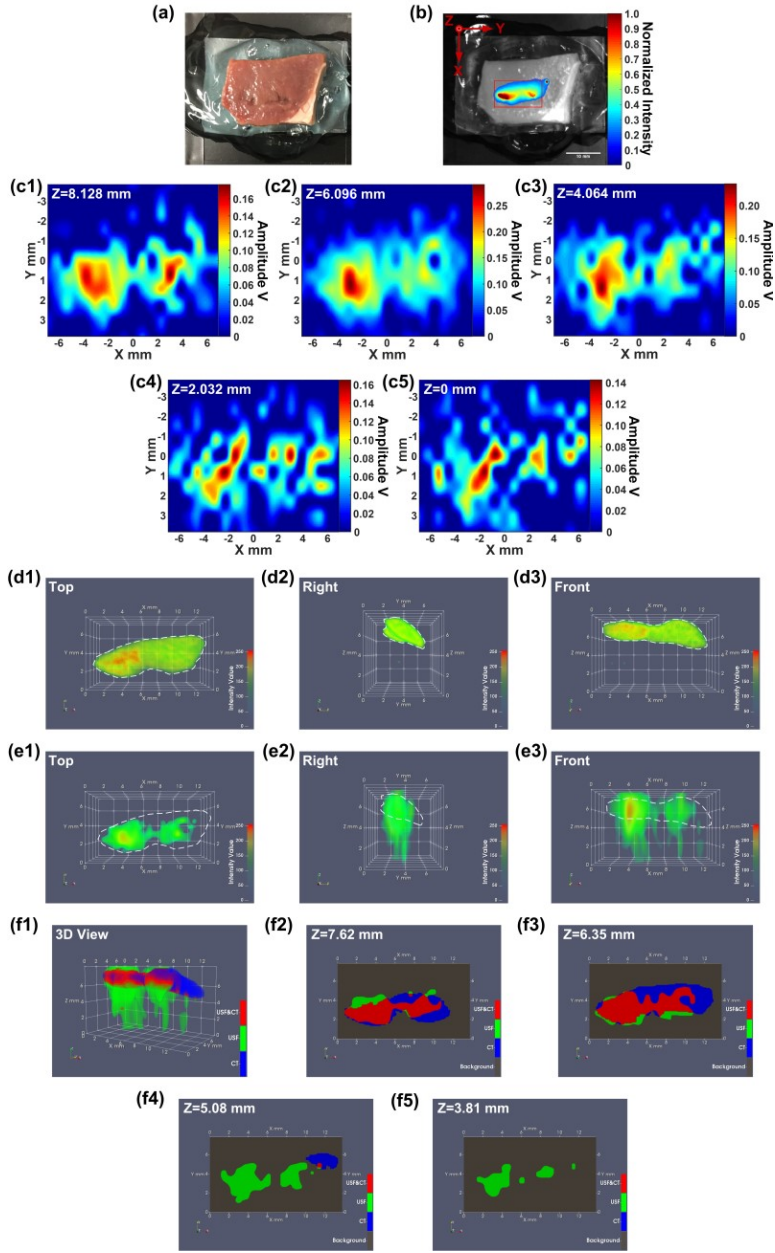

**Figure 6.** (a) A white-light photo of the porcine heart tissue in which the spleen is embedded. (b) The fluorescence image (Ex/Em=808/830 nm) of the tissue and the USF scan area indicated as a red box on the horizontal XY plane. (c1–c5) 2D USF images on XY plane at different depths. (d1–d3) The top view (XY), right view (YZ), and front view (XZ) of the reconstructed 3D CT images. (e1–e3) The top view (XY), right view (YZ), and front view (XZ) of the 3D USF images. (f1) The 3D view of the co-registered image of the two imaging modalities, and (f2–f5) the 2D cross-section images on XY plane at different depths.

## Discussion

**HIFU driving voltage (Vpp):** Figure 1 indicates that raising the HIFU driving voltage will increase the USF signal strength and therefore SNR (Figures 1(b) – (e)) but decrease the spatial resolution (Figures 1(d) and (g)). This is because the higher driving voltage leads to the higher ultrasound power, higher temperature rise, larger thermal focal volume, and therefore more switched-on contrast agents. The driving voltage also varies in different types of tissues due to their different ultrasound absorption abilities. In this study, a total of five different types of tissues were used: (1) tube-embedded porcine muscle tissue, (2) porcine heart tissue, (3) breast tumor in a live mouse, (4) live mouse spleen, and (5) dead mouse spleen. For convenience, the driving voltages applied to these tissues are reorganized in Table S1 in SI. Generally, high driving voltage was adopted in live tissues, such as 200 mV in the live spleen and 250 mV in the live tumor; and relatively low driving voltages in dead tissues, such as 140 mV in the dead spleen, 120 mV in the porcine heart tissue, and as low as 60 mV in the tube-embedded porcine muscle tissue. The live tissues usually contain much more fluids, such as water and blood, while the dead tissues usually have much less. Water and blood usually have much smaller ultrasound absorption coefficients than the tissues mentioned above (water: 0.002; blood: 0.2; muscle: 1.1, breast: 0.75; fat: 0.48; liver: 0.5 db/MHz/cm)[29]. In addition, flowing blood may partially function as a cooling system. Accordingly, live tissues are more difficult to heat via HIFU than dead tissue. This fact can explain why we adopted 200 mV for the *in vivo* and 140 mV for the *ex vivo* spleen imaging when i.v. injection was applied to both, and 250 mV in live tumor and 120 mV in dead porcine heart tissue when local injections were applied to both. Lastly, comparing the tube-embedded porcine muscle tissue with the porcine heart tissue with local injections, even lower voltage can be used for the former. This could be for two possible reasons. (1) The adopted silicone tube has a much larger ultrasound absorption coefficient than that of biological tissues discussed in this study (see Figure S3 in SI). (2) Another factor that should be considered is the agent dilution due to either the local or i.v. injection. In the tube-based phantom, the stock solution of the USF contrast agent was injected into the tube so no dilution is applied. For local injection, the contrast agent concentration is usually diluted due to natural diffusion of agents in tissue, whereas for i.v. injection, the dilution is due to the relatively large blood or body volume. In addition, to avoid HIFU-induced tissue damage, the temperature increases in real tissue are limited to only a few Celsius degrees under the driving voltages used in this study (SI Figure S4).

**Lateral vs. axial resolution:** The lateral resolution of the USF imaging system is ~3.3–4.5 times better than the axial resolution, which is due to the inherent nonuniform shape of the ultrasound focus (Figure 1(d) vs. 1(g)). As shown in all the 3D co-registered images, the sizes of the USF images are larger than CT images along the Z direction, which is the transmission direction of the ultrasound (i.e., the axial direction of the HIFU transducer). However, the USF images match well with CT images on the horizontal XY plane, which is the plane orthogonal to the ultrasound transmission direction as shown in the cross-section images of the XY plane (i.e., the lateral direction of the HIFU transducer).

**Biodistribution:** The biodistribution data indicates that the USF contrast agents are mainly distributed in the spleen and liver. This is because the median hydrodynamic diameter of the nanoparticles is ~335 nm, and the large-sized nanoparticles are more likely to be accumulated in the spleen and liver due to the uptake of macrophages[30]. The two kidneys also show some fluorescence signal, which may be due to the existence of either very small nanoparticles or the residue of free ICG molecules. Currently, it is unclear why the stomach also shows certain fluorescence signal. Similar to the USF agent, the CT contrast agent (ExiTron nano 12000) was mainly found in the spleen and liver. The mean hydrodynamic diameter of the CT contrast agent is ~110 nm, and it mainly accumulates in the liver and spleen due to uptake by macrophages of the liver[31]. The similar biodistribution in the spleen of both contrast agents makes it possible to image this organ via both modalities for comparison.

**Why select spleen for *in vivo* USF imaging:** In this study, we selected the spleen as the target organ because (1) it can accumulate enough contrast agents of both modalities, has a relatively small volume (therefore a relatively high contrast agent concentration), and is close to the skin, which are all helpful to achieve a high SNR. And (2) it has a unique shape and location, which make it easy to be roughly identified and located with a 2D camera. To achieve this goal, we controlled the nanoparticles size relatively large (~335 nm) because large-sized nanoparticles have a better chance of accumulating in the spleen.

**The key factor to successfully achieve *in vivo* USF imaging—Stability of USF contrast agents in biological environments:** This is the first time that USF has been demonstrated to successfully image living biological tissues via both local and i.v. injections. It is important to point out that the success of *in vivo* imaging is highly dependent upon the stability of contrast agents in biological environments. The adopted USF contrast agent in this study is ICG-encapsulated PNIPAM-based nanoparticles. The shelf life of the stock solution can be as long as a year or longer without loss of the switching performance[32]. More importantly, the USF contrast agent adopted in this study shows high stability in biological environments. As described in Methods, the *in vivo* USF imaging of the spleen was conducted between 4 and 9 hours after the i.v. injection, and the *ex vivo* USF imaging was conducted 1 day after the i.v. injection. These data clearly indicated that the switching property of the contrast agents was well maintained no matter whether in living or dead tissue. In addition, we believe that the temperature-switching threshold of the adopted contrast agent was maintained the same in both living and dead spleen as the value measured *in vitro* (such as in a cuvette). This conclusion is drawn based on the following two facts: (1) to be able to observe a similar USF signal in the dead tissue to that in living tissue, the background temperature of the dead tissue needs to be raised up to around 37 °C

(similar to the living mouse body temperature) via a water bath; (2) without heating the dead tissue, almost no USF signal can be observed from the dead tissue (see Figure S8).

**Switching on-to-off ratio of the USF contrast agent:** Currently the on-to-off ratio of the adopted ICG-NPs is ~5, which is enough for *in vivo* USF imaging in this study and may be able to further improve in future by optimizing the synthesis protocols. Using a contrast agent with a higher on-to-off ratio may improve the SNR and therefore possibly reduce the required driving voltage or ultrasound power.

**Current technical limitations and future directions:** The non-uniform spatial resolutions have been partially addressed in our recent work by using two $90^0$-crossed confocal HIFU transducers[25]. Further development and implementation need to be conducted in future. Another technical limitation of current USF imaging systems is the speed. Because it's a point-by-point scanning imaging technology and adopts the HIFU-induced temperature to switch on fluorescence, it usually takes a few hours to finish a 3D image acquisition. Currently, we are developing and testing several strategies to increase the imaging speed, such as via camera-based fast scanning techniques, and we expect the imaging speed to increase significantly if successful. It is highly desirable to develop new USF contrast agents that have better switching performance (such as high on-to-off ratio, narrow-switching bandwidth, in near-infrared region, etc.), are feasible for molecular targeting and multicolor imaging, and are biocompatible and biodegradable. With the success of addressing these technical challenges and approvals of various regulations in the future, we expect that USF imaging technique will be a promising, unique, and powerful technology for preclinical and clinical applications to complement the existing imaging technologies.

## Conclusion

For the first time, we successfully achieved the goal of *in vivo* USF imaging via both local and intravenous injections in mice. The USF contrast agent adopted in this study, ICG-encapsulated PNIPAM nanoparticles, was proved to be stable in biological environments (such as in breast tumor and spleen). Biodistribution studies showed that the contrast agent was mainly accumulated in the spleen of mice. It provided a good opportunity to conduct *in vivo* USF imaging. All the USF images were compared with CT images. The results showed that USF achieved similar accuracy to CT on the lateral plane (i.e., the horizontal XY plane orthogonal to the direction of the ultrasound wave propagation), and lower accuracy than CT in the axial direction (i.e., the direction of the ultrasound wave propagation). In addition, USF maintained high sensitivity and specificity to its contrast agents in deep tissues because the signals could be generated only from the contrast agents, and therefore USF was insensitive to non-specific background photons (or noises). With the success of *in vivo* USF imaging, we believe that many potential biomedical applications can be explored in future.

## Methods

**USF imaging system.** The USF imaging system was similar to the one described in our previous study (Figure S9)[23]. Briefly, the excitation light (808 nm) was modulated by a function generator (FG) at 1kHz and passed through a band-pass filter (785/62) to illuminate the sample. The emitted fluorescence signal was filtered by two long-pass interference filters (830 LP) and two absorptive filters (RG 830) and was received by a photomultiplier (PMT) tube so that the light was converted into an electrical signal. The 1kHz electrical signal was further amplified by a preamplifier and sent to a lock-in amplifier (LIA), which received the reference signal from the FG. The time constant of the LIA played a role as a low-pass filter and was set to 300 ms in all experiments. A HIFU (with a center frequency of 2.5 MHz) was used to heat the sample and switched on the contrast agent in the focus. Another FG was used to generate the HIFU driving signal, and the peak-to-peak voltage of this signal is the driving voltage (Vpp) discussed in the main text. This signal was further amplified by a 50-dB power amplifier (RF-AMP) and sent to a matching network (MNW) and then the HIFU transducer. At the same time, the same FG sent a trigger to the data acquisition card (DAC) to sample the output signals from the preamplifier and the LIA. A motorized translation stage was used to realize sample scanning. The water temperature was controlled by a temperature controller (with a heater and a temperature sensor) and kept uniform by a magnetic stirrer (with a long magnetic bar, 11-100-16S, Fisher Scientific, USA). See Supplementary Information for more details.

**Fluorescence imaging system.** The excitation light from a laser (808 nm, MGL-II-808-2W, Dragon lasers, JL, China) driven by a function generator (FG, 33220A, Agilent, Santa Clara, CA, USA) passed through an excitation filter (FF01-785/62-25, Semrock Inc., Rochester, NY, USA) to illuminate the sample. The emitted fluorescence signal filtered by a set of emission filters (four long-pass filters (two BLP01-830R-50 and two BLP01-830R-25, Semrock Inc., Rochester, NY, USA) and an absorptive filter (FSRRG830, Newport Corporation, Irvine, CA, USA)), passed through a camera lens (35mm fixed-focal-length lens, Edmund Optics Inc., Barrington, NJ, USA) and was received by an electron-multiplying CCD (ProEM®-HS:1024BX3, Princeton Instruments, Trenton, NJ, USA).

**Sample configuration protocol of the silicone tube-embedded porcine muscle tissue phantom and the heart tissue.** A silicone tube (ST 60-011-01, Helix Medical, Carpinteria, CA, USA) was inserted into a piece of porcine muscle tissue with a thickness of 1 cm at a depth of 5 mm from the bottom. To prepare the heart tissue, the USF contrast agent (ICG-NPs) with an LCST of ~24–25 °C was mixed with the commercial CT contrast agent (ExiTron$^{TM}$ nano 12000, Miltenyi Biotec, Bergisch Gladbach, Germany) with a ratio of 3:1 first. The 50μL mixed contrast agent was then locally injected into the heart tissue with a thickness of 1 cm to form a single point. The tissue (heart tissue or porcine muscle tissue) was placed on the transparent parafilm (PM-992, BEMIS Company Inc. Neenah, WI, USA), which sealed the rectangular window of a small tank. The gap between the bottom side of the tissue and the parafilm was filled with ultrasound gel (01-08, Aquasonic 100, Parker Laboratories Inc., Fairfield, NJ, USA) to maintain appropriate ultrasound coupling. The surface of the tissue was also covered with ultrasound gel and a piece of transparent parafilm to prevent it from drying during the experiment (SI, Figure S10).

***In vivo* USF imaging in mouse tumor.** The animal protocols were approved by the University of Texas at Arlington's Animal Care and Use Committee. The mouse (female, 55 weeks) of strain FVB/N-Tg (MMTVneu) 202Mul/J purchased from Jackson Lab (Bar Harbor, ME, USA) was used for this experiment. The mouse was initially anesthetized with a concentration of 2.5% isoflurane (ISOSOL ISOFLURANE, Miller Veterinary Supply, Ft. Worth, TX, USA) at a flow rate of 1 liter/min for animal preparation. The hair on the tumor surface was removed. A temperature image of the whole mouse body was taken by an infrared (IR) camera (FLIR A300, FLIR Systems, Wilsonville, Oregon, USA). Afterwards, the 120 μL mixed contrast agent (90 μL ICG-NPs with an LCST of ~35–36 °C and 30 μL ExiTron nano 12000) was locally injected into the mouse's breast tumor. The mouse then underwent USF imaging via the FD-USF imaging system. The mouse tumor was placed on the parafilm, which was covered by ultrasound gel (SI, Figure S11(b)). To realize long time anesthesia of the mouse, the concentration and the flow rate were reduced to 1.8% and 0.8 liters/min during USF imaging, respectively. To maintain the mouse body temperature, the water temperature in the large tank was kept at 38°C. To determine the scan area, the rough location of the contrast agent was scanned by the USF system quickly without ultrasound exposure. The scan area was selected based on the fluorescence distribution. After USF imaging was completed, the mouse was sacrificed. A 2D planar fluorescence image (excitation/emission: 808/830 nm) was immediately acquired after the mouse's death because the body posture had not been changed at that time. The dead mouse was then scanned by a commercial micro-CT system (Skyscan 1178, Bruker, Kontich, Belgium).

***In vivo* and *ex vivo* USF imaging of mouse spleen.** A BALB/c mouse (female, 7 weeks, 16g) purchased from Jackson Laboratory (Bar Harbor, ME, USA) was used for this experiment. The mouse was anesthetized with 2.5% isoflurane at a flow rate of 1 liter/min initially, and the hair on the whole body surface was removed. ICG-NPs with an LCST of ~35–36 °C and ExiTron nano 12000 were mixed together with a volume ratio of 3:1. The mouse was injected with 150 μL mixed contrast agent via the tail vein, which corresponded to a 153.75 mg/kg dose of ICG-NPs. The size and concentration of the ICG-NPs in its stock solution are 334.8 nm and 24.60 mg/mL, respectively (see SI for details). The 2D planar fluorescence images (excitation/emission: 808/830 nm) of the whole mouse body were taken at 0 and 3 h after the injection. Afterwards, the mouse was placed on the transparent parafilm covered by ultrasound gel with the left side of the body upward for USF imaging (SI, Figure S11(c)). The USF imaging was started at 4 h after the i.v. injection since it took about one hour to prepare the experiment. After the USF imaging and sacrifice of the animal, another 2D planar fluorescence image of the dead mouse was taken at about 9 h after the i.v. injection. The dead mouse was then scanned by the micro-CT system. After CT imaging, the main organs (heart, liver, spleen, kidney, stomach, small intestine, and large intestine) were dissected and put on a parafilm. These *ex vivo* organs were immediately imaged using the 2D planar fluorescence imaging system and the CT imaging system. The spleen was then inserted into a piece of porcine heart tissue with a thickness of 1 cm for *ex vivo* USF imaging. The spleen was placed near the top of the heart tissue to simulate the situation when the spleen was in the mouse body. The 2D planar fluorescence imaging, USF imaging, and CT imaging of the heart tissue with spleen inside were carried out in sequence. To realize *ex vivo* USF imaging, the water temperature was kept at 37 °C.

**Image processing and co-registration.** Briefly, a raw USF signal was recorded at each position on the scan plane as a function of time. By fitting the raw USF signal and finding the maximum value in the time window of 2 to 5 s at each location, the USF signal strength at each location was extracted. Similar to fluorescence confocal or 2-photon microscopy, USF adopted a point-by-point scanning method. Therefore the USF signal strength was directly correlated to the distribution of the contrast agent, although it was not exactly equal. Consequently, in this study the spatial distribution of the USF signal strengths was used to represent the contrast agent distribution. Stricter methods can be seen in the section of USF image processing in SI. The noise and artifacts were removed by correlating the fitted USF signal with three typical reference signals. In each experiment, all fitted USF data were evaluated, and three typical USF curves were selected. Then, the selected three data were fitted, and the resultants were used as the reference signals for correlation (see Figure S5(b)). The quality of the USF image was further optimized by image segmentation via the 2D Otsu method[33]. The CT image was reconstructed and processed via the software provided by the manufacturer of the micro-CT (NRecon and CTAn). The scanning volume in USF imaging was extracted from the reconstructed CT image by comparing the 2D planar fluorescence image with the X-ray projection image of the phantom or the mouse. The 3D USF image and the 3D CT image were co-registered in the software provided by the manufacturer of the micro-CT (DataViewer). Because the rough locations of the contrast distribution shown in both imaging modalities were very close to each other, the co-registration was realized by slightly adjusting the angle and location of the objects. The visualization of the three-dimensional images was realized via ParaView (Sandia National Laboratory, Kitware Inc, Los Alamos National Laboratory). The details of the USF image processing are shown in the section on USF image processing in SI.

**Acknowledgements:**

This work was supported in part by funding from the CPRIT RP170564 (Baohong Yuan) and the NSF CBET-1253199 (Baohong Yuan). Thanks for Dr. Kytai Nguyen and Dr. Yi Hong for allowing us to use some equipment in their lab to synthesize and characterize the contrast agent.

**Author Contributions**

B.Y. was the principal investigator and conceived the idea. T.Y. and B.Y. developed the idea and designed the experiments. T.Y. was the main operator of the experiments. T.Y. and S.Y. conducted the *in vivo* USF imaging of mouse spleen. S. Y. and Y. L. synthesized and characterized the contrast agents supervised by B.Y. T.Y. and B.Y. explained the data and prepared the manuscript. All authors reviewed the manuscript.

**Additional Information**

Supplementary information accompanies this paper
Competing financial interests: The authors declare no competing financial interests.